\documentclass[aps,prb,onecolumn,showpacs,showkeys]{revtex4}
\usepackage{epsfig}
\usepackage{amsmath}
\begin{document}

\title{Zero-conductance resonances and spin filtering effects in ring conductors
subject to Rashba coupling}
\author{R. Citro, F. Romeo and M. Marinaro}
\affiliation{Dipartimento di Fisica ``E. R. Caianiello",
Universit{\`a} degli Studi di Salerno, and Unit{\`a} C.N.I.S.M.
\\ Via S. Allende, I-84081 Baronissi (SA), Italy }

\date{\today}

% ----------------------------------------------------------------
\begin{abstract}
We investigate the effect of Rashba spin-orbit coupling and of a
tunnel barrier on the zero conductance resonances appearing in a
one-dimensional conducting Aharonov-Bohm (AB) ring symmetrically
coupled to two leads. The transmission function of the
corresponding one-electron problem is derived within the
scattering matrix approach and analyzed in the complex energy
plane with focus on the role of the tunnel barrier strength on the
zero-pole structure characteristic of transmission
(anti)resonances. The lifting of the real conductance zeros is
related to the breaking of the spin-reversal symmetry and
time-reversal symmetry of Aharonov-Casher (AC)and AB rings, as
well as to rotational symmetry breaking in presence of a tunnel
barrier. We show that the polarization direction of transmitted
electrons can be controlled via the tunnel barrier strength and
discuss a novel spin-filtering design in one-dimensional rings
with tunable spin-orbit interaction.
\end{abstract}


\maketitle

\section{Introduction}
In the last decade enormous attention, from both experimental and
theoretical physics communities, has been devoted towards control
and engineering of the spin degree of freedom at the mesoscopic
scale, usually referred to as spintronics.\cite{spintr_1,spintr_2}
The major goal in this field is the generation of spin-polarized
currents and their appropriate manipulation in a controllable
environment, preferably in semiconductor systems. Since the
original proposal of the spin field effect transistor (spin FET)
by Datta and Das\cite{FET_datta}, many proposals have appeared,
but the realization of a spin transistor or spin filter still
remains challenging. Of particular interest appear setups based on
intrinsic spin dependent properties of semiconductors, like the
Rashba spin orbit (SO) effect\cite{rashba_1,rashba_2} for a
two-dimensional electron gas confined to an asymmetric potential
well. This is a dominant mechanism for the spin-splitting in
semiconductors that has been proven to be a convenient means of
all-electrical control of spin polarized current through
additional gate voltages\cite{nitta_electric_contr}. In addition,
suitable means for controlling spin at mesoscopic scales are
provided by quantum interference effects in coherent ring
conductors under the influence of electromagnetic potentials,
known as Aharonov-Bohm\cite{aharonov_bohm} and
Aharonov-Casher\cite{aharonov_casher} effect. This possibility has
driven a wide interest in spin-dependent Aharonov-Bohm physics,
and the transmission properties of mesoscopic AB and AC rings
coupled to current leads have been studied under various aspects
such as AB flux and coupling dependence of
resonances\cite{buttiker_ring,sigrist_as}, Berry
phases\cite{berry_original,loss_ring,stern_ring,aronov_ring,qian_ring,hentschel_ring},
spin-related conductance
modulation\cite{nitta_ring_so,frustaglia_ring_so,chao_ring_so},
persistent currents\cite{loss_ring,governale_ring}, spin
filters\cite{spin_filter_ring,foldi_05} and
detectors\cite{detector_spin_ring}, spin
rotation\cite{rotation_spin_ring}, and spin switching
mechanisms\cite{switching_spin_ring}.

In this paper we focus on the zero conductance resonances in a
one-dimensional AB ring subject to Rashba spin-orbit
interaction\cite{note_dresselhaus} and interrupted by a tunnel
barrier in the lower arm. First, motivated by the work of
Aeberhard et al.\cite{sigrist_as}, we revisit the subject of
spin-induced modulation and of zero-pole structure of the
conductance of a symmetrically coupled ring as a function of the
Rashba coupling strength and extract distinct effects due to the
presence of tunnel barrier in one of the arms which have not been
considered in earlier works. Using the effective Hamiltonian for
one-dimensional (1D) rings, as recently considered in
Refs.\cite{frustaglia_ring_so,molnar_ring_so} and taking into
account the corresponding appropriate eigenstates, in Sec.II we
derive the analytic expression of the conductance in presence of
AB-AC fluxes and of the tunnel barrier. Later on, we discuss the
real zeros conductance and the zeros lifting due to the SO
interaction and the tunnel barrier strength. The imprints of the
presence of a tunable tunnel barrier together with the Rashba
coupling (strength) on the overall conductance is remarkable. In
fact, in Sec.III we demonstrate, by means of numerical and
analytical calculations, that the interference zeros of the
spin-resolved conductance in one spin-channel can be compensated
by poles, while the location of zeros due to the presence of the
tunnel barrier can be controlled by means of its strength. This
implies a spin-filter mechanism which is probably more convenient
for experimental realizations than previous
proposals\cite{switching_spin_ring,spin_filter_ring,shen_spin_precession},
due to presence of intrinsic effective field associated to the
Rashba interaction and to a highly controllable parameter, the
tunnel barrier strength. We present a short summary in
Sec.\ref{sec:concl}.

\section{The AB ring with tunable Rashba SO coupling}

\subsection{Hamiltonian and the one-particle solution}

As known, spin-orbit coupling (SOC) is due to a magnetic field
generated in the reference frame of the moving electron by an
electric field in the reference frame of the laboratory. When
considering a one dimensional ring in a semiconductor structure,
an effective Rashba electric field results from the asymmetric
confinement along the direction ($k$) perpendicular to the plane
of the ring\cite{rashba_1,rashba_2}. A consequence of lack of
inversion symmetry in presence of a confinement potential $V(k)$
is a spin band splitting proportional to the momentum of the
electron. The Hamiltonian describing SO coupling is the following:
\begin{equation}
\hat{H}_{SO}=\frac \alpha \hbar
(\hat{\overrightarrow{\sigma}}\times \hat{\overrightarrow{p}})_k,
\end{equation}
where $\hbar/2 \hat{\overrightarrow{\sigma}}$ is the spin operator
expressed in terms of the Pauli spin matrices,
$\hat{\overrightarrow{\sigma}}=(\sigma_i,\sigma_j,\sigma_k)$  and
$\alpha$ is the SOC associated to the effective electric field
along the $k$ direction. The total Hamiltonian of a moving
electron in presence of SOC can be found in
Ref.[\onlinecite{meijer_ham}]. In the case of a one-dimensional
ring an additional confining potential, $V_c(r)$ must be added in
order to force the electron wave function to be localized on the
ring. A typical confining potential is an harmonic potential
centered at the radius of the ring, $V_c(r)=\frac 1 2 K (r-R)^2$.
When only the lowest radial model is taken into account, the
resulting effective one dimensional Hamiltonian in a dimensionless
form\cite{frustaglia_ring_so,molnar_ring_so} can be written as:
\begin{equation}
\label{eq:ham_1d} {\hat H}=\frac{2m^\star
R^2}{\hbar^2}\hat{H}_{1D}=\left( -i \frac{\partial}{\partial
\varphi}+\frac{\beta}{2}\sigma_r\right)^2,
\end{equation}
where $m^\star$ is the effective mass of the carrier,
$\beta=2\alpha m^\star/\hbar^2$ is the dimensionless SOC,
$\sigma_r=\cos \varphi \sigma_i+\sin \varphi \sigma_j$, and
additional constants have been dropped. The parameter $\alpha$
represents the average electric field along the $k$ direction and
is assumed to be a tunable quantity. For an InGaAs-based
two-dimensional electron gas, $\alpha$ can be controlled by a gate
voltage with typical values in the range $(0.5\div 2.0)\times
10^{-11}$eVm\cite{param1,param2}. In presence of a finite AB
magnetic flux and of a tunnel barrier localized in the lower arm
of the ring, the Hamiltonian can be generalized to:
\begin{equation}
\label{eq:ham_ac_ab} {\hat H}=\left( -i \frac{\partial}{\partial
\varphi}+\frac{\beta}{2}\sigma_r-
\frac{\Phi_{AB}}{\phi_0}\right)^2+v\delta(\varphi'+\frac{\pi}{2}),
\end{equation}
where $\phi_0=h c /e$ is the quantum flux  and $v$ is the
dimensionless tunnel barrier strength $v=2m^\star R^2 V/\hbar^2$
which can be tuned by an external gate voltage applied at a
quantum point contact\cite{ring_exp} and $\varphi'=-\varphi$. As
outlined in the Appendix of Ref.[\onlinecite{molnar_ring_so}] one
can solve the eigenvalue problem in a straightforward manner and
the energy eigenvalues are:
\begin{equation}
E^\sigma_n=(n-\Phi_{AC}^\sigma/2\pi-\Phi_{AB}/2\pi)^2,
\end{equation}
where $\sigma=\pm$, $\Phi_{AC}^\sigma$ is the so-called
Aharonov-Casher phase\cite{aharonov_casher}:
\begin{equation}
\Phi_{AC}^\sigma=-\pi(1-\sigma \sqrt{\beta^2+1}).
\end{equation}
At fixed energy, the dispersion relation yields the quantum
numbers $n_\lambda^\sigma (E)=\lambda \sqrt{E}+\Phi^\sigma/2\pi$,
where we have introduced $\Phi^\sigma=\Phi_{AC}^\sigma+\Phi_{AB}$,
and the index $\lambda=\pm$ refers to right/left movers,
respectively. The eigenvectors have the general
form\cite{frustaglia_ring_so,molnar_ring_so}:
\begin{equation}
\Psi^\sigma_n(\varphi)=e^{in\varphi}\chi^\sigma(\varphi),
\end{equation}
where $n\in$Z is the orbital quantum number.  It should be noted
that the spinors $\chi^\sigma(\varphi)$ are generally not aligned
with the Rashba electric field, but they form a tilt angle given
by $\tan \theta=-\beta$ relative to the $k$ direction. The
mutually orthogonal spinors $\chi^\sigma(\varphi)$ can be
expressed in terms of the eigenvectors $\begin{pmatrix}
  1 \\
  0
\end{pmatrix}$,
$\begin{pmatrix}
  0 \\
  1
\end{pmatrix}$, of the Pauli matrix $\sigma_k$, as
\begin{equation}
\chi^{\sigma=+}(\varphi)=\frac{1}{\sqrt{2\pi}}\begin{pmatrix}
  \cos \theta/2 \\
  e^{i\varphi} \sin \theta/2
\end{pmatrix}
\end{equation},
\begin{equation}
\chi^{\sigma=-}(\varphi)=\frac{1}{\sqrt{2\pi}}\begin{pmatrix}
  \sin \theta/2 \\
  -e^{i\varphi} \cos \theta/2
\end{pmatrix}
\end{equation}.

\subsection{Device geometry, boundary conditions and scattering matrix}

\begin{figure}[htbp]
\centering
\includegraphics[scale=0.5]{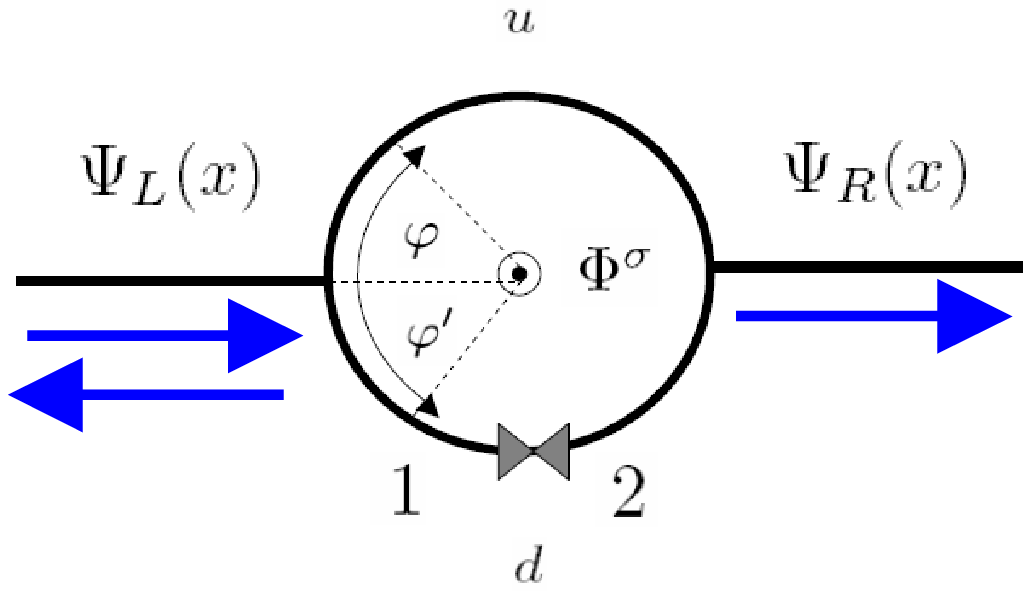}\\
\caption{The one-dimensional ring symmetrically coupled to
conducting leads and interrupted by a tunnel barrier in the lower
arm.} \label{fig:device}
\end{figure}

The ring connected to the two leads and interrupted by the tunnel
barrier in lower arm is shown in Fig.\ref{fig:device}. If the ring
is not connected to external leads the boundary condition on the
wave function implies that it has to be single valued when the
argument $\varphi$ is increased of $2\pi$, which gives integer
quantum numbers. When the ring is connected to external leads the
boundary conditions are altered. In this case it is appropriate to
apply the spin-dependent version of the Griffith boundary's
condition\cite{griffith_boundary} at the intersection. This
reduces the description of the electron transport to a
one-dimensional scattering problem. These boundary conditions
state that (i) the wave function must be continuous and (ii) the
spin density must be conserved. The same conditions on the
continuity of the wave function and on the density conservation
apply at the position of the tunnel barrier in the lower arm.
Below, we briefly outline the determination of transmission and
reflection coefficients of the scattering matrix problem (details
can be found, e.g. in Ref.\onlinecite{molnar_ring_so}).

 Assumed that electrons in the two leads are free and
have momentum $k$, the corresponding energy is $\hbar^2
k^2/2m^\star$. When an electron moves along the upper arm in the
clockwise direction from the input intersection at $\varphi=0$
(see Fig.\ref{fig:device}), it acquires a phase $\Phi^\sigma/2$ at
the output intersection $\varphi=\pi$, whereas the electron
acquires a phase $-\Phi^\sigma/4$ in the counterclockwise
direction along the other arm when moving from $\varphi'=0$ to
$\varphi'=\pi/2$ and from $\varphi'=\pi/2$ to $\varphi'=\pi$,
respectively. Therefore the total phase is $\Phi^\sigma$ when the
electron goes through the loop. The electric field in the ring
changes the momenta of the electrons in different spin states
$\chi^\sigma$ as $k_+^\sigma=k+\Phi^\sigma/2\pi R$ and
$k_-^\sigma=k-\Phi^\sigma/2\pi R$, where the subscript $\pm$
denotes the chirality. The wave functions in the upper($u$) and
lower($d$) arm of the ring can be written as:
\begin{eqnarray}
\Psi_{u}(\varphi)=\sum_{\sigma=\pm,\lambda=\pm}
c_{u,\sigma}^\lambda e^{i n_\lambda^\sigma \varphi}
\chi^\sigma(\varphi), \nonumber \\
\Psi_{d\alpha}(\varphi')=\sum_{\sigma=\pm,\lambda=\pm}
c_{d\alpha,\sigma}^\lambda e^{-in_\lambda^\sigma
\varphi'}\chi^\sigma(\varphi'),
\end{eqnarray}
where the index $d\alpha=d1,d2$ denotes the wave function in the
two-halves of the lower branch and $n_\lambda^\sigma=\lambda k
R+\Phi^\sigma/2\pi$. The wave function of the electron incident
from the left lead in the left and right electrodes can be
expanded as:
\begin{equation}
\Psi_L(x)=\Psi_i+(r_\uparrow,r_\downarrow)^Te^{-ik x},\mbox{
}\Psi_R(x)=(t_\uparrow,t_\downarrow)^T e^{ik x},
\end{equation}
where $x=R \varphi$, $r_\sigma$ and $t_\sigma$ are the
spin-dependent reflection and transmission coefficient, $\Psi_i$
is the wave function of the injected electron $\Psi_{i}=e^{ik
x}\chi^\sigma(0)$. For an incident electron from the right lead an
analogous expansion in terms of reflection and transmission
coefficients is possible with ${i_\sigma,r_\sigma}$ (for left
lead) and ${t_\sigma,0}$ (for right lead) replaced by
${0,t'_\sigma}$ and ${r'_\sigma,i'_\sigma}$. This enables us to
formulate the scattering matrix equation of the ring system as
$\hat{o}=\hat{S}\hat{i}$, where $\hat{o},\hat{i}$ stand for
outgoing and incoming wave coefficients. In particular, the
following relation holds:
$t_\sigma=\sum_{\sigma'}T_{\sigma\sigma'}i_{\sigma'}$,
$r_\sigma=\sum_{\sigma'}R_{\sigma\sigma'}i_{\sigma'}$ and a
similar one for $t'_\sigma,r'_\sigma$. Note that since the
transmission amplitude does not depend on the choice of spinor
basis, i.e., it is invariant under spin rotation, hereafter we
make use of the ring-spinor basis where the spin flip amplitudes
vanish, and $T_{\sigma,\sigma'}$ is diagonal, thus we define
$T_{\sigma,\sigma}=T_\sigma$ and similarly for $R_\sigma$. The
previous expression for wave functions are now used to calculate
the transmission amplitude for the ring system from the proper
requirements on wave function continuity and probability current
conservation. The first Griffith boundary
condition\cite{griffith_boundary} states that the wave function is
continuous at $\varphi=0,\varphi'=0$ and
$\varphi=\pi,\varphi'=\pi$. Concerning the second boundary
condition, if one assumes that there are no spin-flip-processes at
the junctions, it requires that the spin current $J^\sigma=
\text{Re} \lbrack (\Psi^\sigma \chi^\sigma )^\dagger \left( -i
\frac{\partial}{\partial
\varphi}+\frac{\beta}{2}\sigma_r-\frac{\Phi_{AB}}{\phi_0}\right)(\Psi^\sigma
\chi^\sigma )\rbrack$ for each spin direction should be conserved,
i.e. $J^\sigma_u+J^\sigma_d+J^\sigma_{L(R)}=0$. At the point
$\varphi'=\pi /2$ in the lower branch the same boundary conditions
apply in presence of the delta tunnel barrier. Thus the system of
equations to be solved is the following:
\begin{eqnarray}
&&1+r_{\sigma}=c^{+}_{u,\sigma}+c^{-}_{u,\sigma} e^{-i
\frac{\Phi^{\sigma}}{2}}=c^{+}_{d2,\sigma}+c^{-}_{d2,\sigma} e^{i
\frac{\Phi^{\sigma}}{4}}\\
&&c^{+}_{d2,\sigma} e^{i \frac{\phi-\Phi^{\sigma}}{4}}+
c^{-}_{d2,\sigma}e^{-i
\frac{\phi}{4}}=c^{+}_{d1,\sigma} e^{i \frac{\phi}{4}}+ c^{-}_{d1,\sigma} e^{i \frac{\Phi^{\sigma}-\phi}{4}}\\
&&c^{+}_{u,\sigma} e^{i \frac{\phi+\Phi^{\sigma}}{2}}+
c^{-}_{u,\sigma} e^{-i
\frac{\phi}{2}}=c^{+}_{d1,\sigma} e^{i \frac{2\phi-\Phi^{\sigma}}{4}}+ c^{-}_{d1,\sigma} e^{-i \frac{\phi}{2}}=t_{\sigma} e^{i \frac{\phi}{2}}\\
&&1-r_{\sigma}=c^{+}_{u,\sigma}-c^{-}_{u,\sigma}e^{-i \frac{\Phi^{\sigma}}{2}}+c^{+}_{d2,\sigma}-c^{-}_{d2,\sigma} e^{i \frac{\Phi^{\sigma}}{4}}\\
&&c^{+}_{d2,\sigma}e^{i\frac{\phi-\Phi^{\sigma}}{4}}-c^{-}_{d2,\sigma}
e^{-i \frac{\phi}{4}}-c^{+}_{d1,\sigma} e^{i
\frac{\phi}{4}}+c^{-}_{d1,\sigma} e^{i
\frac{\Phi^{\sigma}-\phi}{4}}=
-i z (c^{+}_{d1,\sigma} e^{i \frac{\phi}{4}}+c^{-}_{d1,\sigma} e^{i \frac{\Phi^{\sigma}-\phi}{4}})\\
&&c^{+}_{u,\sigma}e^{i
\frac{\phi+\Phi^{\sigma}}{2}}-c^{-}_{u,\sigma}e^{-i
\frac{\phi}{2}}+c^{+}_{d1,\sigma} e^{i
\frac{2\phi-\Phi^{\sigma}}{4}}-c^{-}_{d1,\sigma} e^{-i
\frac{\phi}{2}}=t_{\sigma}e^{i \frac{\phi}{2}},
\end{eqnarray}
where we have introduced $\phi=2\pi k R=2k L$. In the limit of
zero tunnel barrier in the lower branch the above equations reduce
to those in Ref.\onlinecite{molnar_ring_so}.
 After some algebra we
obtain the transmission coefficients
$t_\sigma(\phi,\Phi^\sigma,z)$ and $t'_\sigma(\phi,\Phi^\sigma,z)$
where $z=v/k$.

The explicit expression of transmission coefficient
$t_{\sigma}(\phi,\Phi^\sigma,z)$, written in a compact form, is
the following:
\begin{eqnarray}
\label{eq:trans} t_{\sigma}(\phi,\Phi^\sigma,z)=\frac{8\,\sin
(\frac{\phi}{4})\left( -4\,\cos (\frac{\phi}{4})\,\cos
(\frac{\Phi^{\sigma}}{2} ) + z\,\sin
(\frac{\phi}{4})\,e^{i\frac{\Phi^{\sigma}}{2}}\right)}
    {4\,z\,\cos (\frac \phi 2) - 2\,\left( 5\,i + 2\,z \right) \,\cos (\phi) +
    i\,\left( 2 + 8\,\cos (\Phi^{\sigma} ) - 2\,z\,\sin (\frac \phi 2) + \left( 8\,i + 5\,z \right) \,\sin (\phi) \right)}.
\end{eqnarray}
Its expression is determined by the tunnel barrier strength, the
total phase, the kinetic state of the incident electrons, the
electric field and the magnetic flux.  In the limit $z\rightarrow
0$, it reduces to\cite{molnar_ring_so}:
\begin{eqnarray}
\label{eq:trans_z0} t_{\sigma}(\phi,
\Phi^\sigma)=\frac{8i\,\cos(\frac{\Phi^\sigma}{2})\sin ({\phi\over
2})}{1 - 5\,\cos (\phi) + 4\,\cos (\Phi^{\sigma}) + 4\,i\,\sin
(\phi)}.
\end{eqnarray}

The conductance in the mesoscopic structure under consideration
can be expressed by means of the Landauer-B\"{u}ttiker conductance
formula\cite{buttiker_cond,landauer_cond}, which in our case
reads:
\begin{equation}
G=(e^2/h)\sum_{\sigma=\uparrow,\downarrow}|T_\sigma|^2=G_\uparrow+G_\downarrow,
\end{equation}
where $T_\sigma$ is the (spin dependent) transmission
amplitude\cite{note} introduced above, and $G_\sigma$ is the
spin-resolved conductance.

Already at this point one can envisage an application of the
device as a spin filter. Assuming one can tune the phases
$\Phi_{AB}$ and $\Phi_{AC}$ (via the magnetic field and the Rashba
strength $\beta$) independently, and varying the tunnel barrier
strength, one can make the ring almost transparent with high
transmission probability only for electrons with spin quantum
number $\sigma=\uparrow(\downarrow)$ and totally opaque to the
electrons with $\sigma=\downarrow (\uparrow)$ by requiring
$|T_\downarrow(\uparrow)|^2=0$ and
$|T_\uparrow(\downarrow)|^2\ne0$.
%The first is the
%finite transmission probability in the spin channel opposite to
%the incident spin orientation. This is the result of spin
%precession along the ring branches due to SOC as considered in
%Ref. 32. The conductance zeros in the opposite channel correspond
%to a frequency of precession which reproduces the incident spin
%orientation at the right junction.
This effect will be discussed in details in next sections where we
will focus on the effect of the tunnel barrier on spin-polarized
transport.

\section{Transmission amplitude from one-electron scattering
formalism}

\subsection{General features}

In quasi-one-dimensional (1D) systems, real conductance zeros
appear under the condition of conserved time reversal
symmetry\cite{lee_99,lee_01} (TRS). The (anti)resonances in the
transmission due to local quasibound states correspond to a
specific zero-pole structure in the complex energy
plane\cite{porod1,porod2,deo,price}. The application of an
external magnetic field modifies this zero-pole structure,
shifting the transmission zeros away from the real axis, with the
shift as a function of the AB phase\cite{kim_02}. Thus, the
lifting of zeros is related to the breaking of TRS. In analogy
with the time-reversal symmetry breaking in AB ring, the lifting
of the real conductance zeros  in an AC ring can be related to
spin-parity symmetry breaking as shown in
Ref.\onlinecite{sigrist_as}. In the following we discuss the
effect of both TRS and spin-parity breaking, together with a
rotational symmetry breaking in the presence of a tunnel barrier
in the lower arm of the ring and will describe the novel
fingerprints associated to the tunable barrier.

\subsubsection{zero tunnel barrier}

\begin{figure}[htbp]
\centering
\includegraphics[scale=0.5]{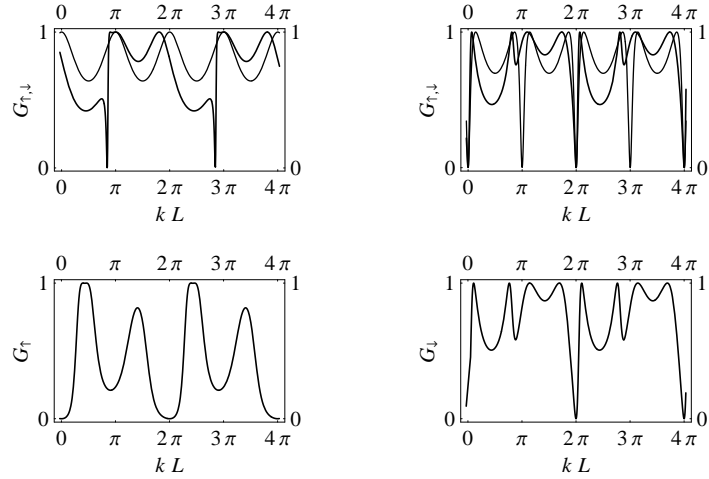}\\
\caption{The spin-resolved conductance $G_\sigma$ as a function of
$k L$. Left upper panel: $\frac{\Phi^\sigma}{2\pi}=0$, $z=0$ (thin
line) and $z=1$ (thick line). Right upper panel:
$\frac{\Phi^\sigma}{2\pi}=0.12$ (with
$\frac{\Phi_{AB}}{2\pi}=0.12,\beta=0$), $z=0$ (thin line) and
$z=1$ (thick line). When $z\ne 0$ periodic transmission zeros
appear on top of the interference zeros at the nodes. Down panels:
$\frac{\Phi_{AB}}{2\pi}=0.12,\beta=1.2$, $z=1$. In the presence of
SOC the two spin channels are distinct and real zeros are lifted.
}\label{fig:trans_vs_z}
\end{figure}

In absence of the tunnel barrier for SOC $\beta=0$ and zero
magnetic flux $\Phi_{AB}$,  the transmission function in
Eq.(\ref{eq:trans}) displays a peculiar resonant behavior as shown
in  Fig.\ref{fig:trans_vs_z}. The oscillation in the conductance
for $z=0$ is due to imperfect transmission caused by the coupling
of the ring and the leads and therefore leads to resonances as a
consequence of backscattering effects\cite{buttiker_ring}. In the
case $\beta\ne 0$ and for non-zero magnetic flux, i.e. for total
flux $\Phi^\sigma\ne 0$, the transmission zeros are obtained from
Eq.(\ref{eq:trans_z0}) as the solution of the equation:
\begin{equation}
\sin(\frac{\phi}{2})\cos(\frac{\Phi^\sigma}{2})=0,
\end{equation}
i.e. $\phi_{0,1}=2 n\pi$, $n\in$ Z. Such zeros correspond to the
interference condition at the nodes. Besides, when $\phi/2\pi$ is
not an integer, zeros can be obtained for $\Phi^\sigma=(2n+1)\pi$,
$n\in$Z.
\subsubsection{non-zero tunnel barrier}
When $z\ne 0$, and in presence of a net total flux, two types of
zero appear. The zeros $\phi_{0,1}$, and the zeros $\phi_{0,2}$
which are determined by the geometry dependent interference
condition at the tunnel barrier. From Eq.(\ref{eq:trans}) the
structural zeros determined by the tunnel barrier are given by the
solution of the equation:
\begin{equation}
\label{eq:zeros2} z \tan(\frac{\phi}{4})
e^{i\frac{\Phi^\sigma}{2}}=4\cos \frac{\Phi^\sigma}{2}.
\end{equation}
This equation, which is satisfied by values of the total flux
$\Phi^\sigma=2 n \pi, n\in$ Z, gives the zeros of second type:
\begin{equation}
\label{eq:zero_2} \frac{\phi_{0,2}}{2}=(2n+1)\pi-2 \arctan(\frac z
4).
\end{equation}
%For $\beta=0$ and $\Phi^{AB}=0$, the phase factor
%$e^{i\frac{\Phi^\sigma}{2}}$ equals unity and the
%Eq.(\ref{eq:zeros2}) simplifies to
%\begin{equation}
%\label{eq:zeros22} \pm \frac{z}{4}=\cot \frac{\phi}{4}.
%\end{equation}
%In the limit $z\rightarrow 0$ the zeros of second type disappear
%and the only zeros are those corresponding to the eigenstates of
%the closed ring.
%In the limit $z\rightarrow 0$, the poles related to the
%transmission resonances are determined by:
%\begin{equation}
%1-5 \cos \phi+4 \cos \phi^\sigma+4i \sin \phi=0.
%\end{equation}
The results for the spin-dependent transmission probability zeros
are summarized in Fig.\ref{fig:trans_vs_z}. In absence of external
electromagnetic fluxes periodical transmission zeros start to
appear when $z\ne 0$, as discussed above.

\subsubsection{conductance zeros and transmission resonances}

By examination of the transmission amplitude in the complex energy
plane we find a certain relation between the conductance zeros and
the transmission resonances.  To examine this connection, it is
convenient to analyze the transmission amplitude in the complex
energy plane by making the substitution $k\rightarrow k_R+i k_I$
and defining $x=e^{-k_IL}e^{ik_RL}$. In terms of the variable $x$
the transmission amplitude can be rewritten as:
\begin{eqnarray}
\label{eq:trans_x} T_\sigma(x,\Phi^\sigma,z)=\frac{-4e^{i
\frac{\Phi^\sigma}{2}}(x-1)\{x[(z-2i)e^{i\Phi^{\sigma}}-2i]-(z+2i)e^{i\Phi^{\sigma}}-2i\}}
{x^{4}(z-2i)e^{i\Phi^{\sigma}}+x^{3}(2ze^{i\Phi^{\sigma}})+2ix^{2}(4e^{i2\Phi^{\sigma}}+2e^{i\Phi^{\sigma}}+4)+x(6z
e^{i\Phi^{\sigma}})-9(z+2i)e^{i\Phi^{\sigma}}}.
\end{eqnarray}
%The physics of the system
%is affected by the zero-poles in the vicinity of the unitary
%circunference determined by the condition $x=e^{-k_IL}e^{ik_RL}=1$
The real zeros conductance are given by points belonging to the
unitary circumference $|x|^2=1$ (which correspond to $k_I=0$),
whereas the poles have a finite imaginary part and correspond to
points away from the unitary circumference.
%The real part of the energies of the
%zeros and poles are not identical which results in an asymmetric
%shape of the resonance (Fano types)\cite{31}.
The poles correspond to maximum in the conductance curve, or to
resonances (Fano types)\cite{fano} when $k_I$ is finite but small.
To find the real zeros conductance, we first solve the numerator
w.r.t. $x$ and then impose the condition $|x|^2=1$. As in the
analysis above, when $z$ is nonzero we can distinguish two types
of zeros: Those corresponding to the interference condition at the
nodes for $x=1$, i.e. $\phi_{0,1}=2 k_R L=2\pi n$ ($n$ even) and
those determined by the scattering at the tunnel barrier for:
%From the
%condition $|x|^2=1$:
%\begin{eqnarray}
%|x|^2={z^2+8(1+\cos(\Phi^{\sigma}))+4z\sin(\Phi^{\sigma})\over
%z^2+8(1+\cos(\Phi^{\sigma}))-4z\sin(\Phi^{\sigma})},
%\end{eqnarray}
%From Eq.(\ref{eq:trans_x}), it is immediate to see that in the
%limit $z\rightarrow 0$ only real-zeros conductance of first type
%are obtained, whereas in the opposite limit, $z\rightarrow \infty$
%the real-zeros conductance are given by $\phi_\infty=2n\pi,
%n\in$Z. When $z$ is finite, we first solve the equation in the
%numerator of Eq.(\ref{eq:trans_x}) w.r.t. $x$:
\begin{equation}
\label{eq:zeros_complex} x={z
e^{i\Phi^{\sigma}}+2i(1+e^{i\Phi^{\sigma}})\over z
e^{i\Phi^{\sigma}}-2i(1+e^{i\Phi^{\sigma}})}.
\end{equation}
Imposing the condition $|x|^2=1$, which is satisfied by values of
the total flux $\Phi^\sigma=2 n \pi, n\in$ Z, the zeros of second
type are those given by Eq.(\ref{eq:zero_2}). Their expression
implies that for integer total flux the position of the real zeros
conductance on the unitary circumference can be controlled by $z$.

\begin{figure}[htbp]
\centering
\includegraphics[scale=0.5]{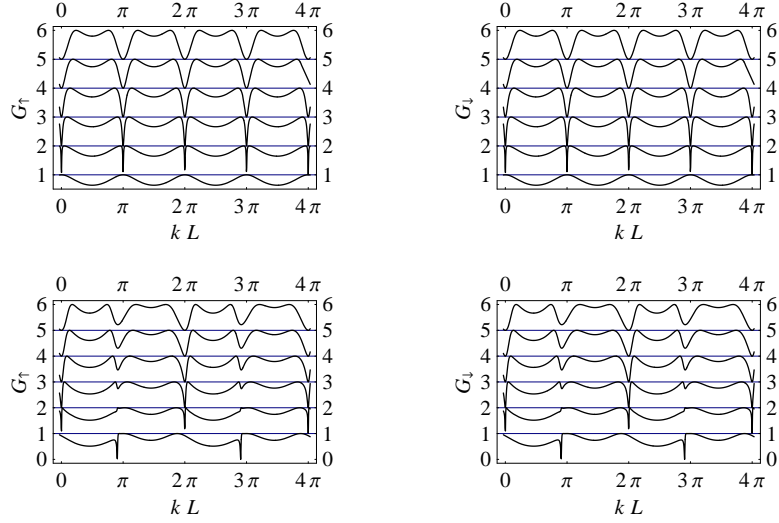}\\
\caption[Fig.3]{The spin resolved conductance $G_\sigma$ versus $k
L$ for different values of the Aharonov-Bohm flux
$\frac{\Phi_{AB}}{2\pi}=0.04\times n$, where $n=0,1,2,3,4,5$ from
bottom to top and $z=0$ (upper panels), $z=0.6$ (lower panels).
The SOC has been fixed at $\beta=0$. In the upper panels only
interference zeros appear ($kL=n\pi$), in the lower panel
structural zeros conductance appear for $z\ne 0$ as discussed in
the text. }\label{fig:effect_z_cond}
\end{figure}

\begin{figure}[htbp]
\centering
\includegraphics[scale=0.5]{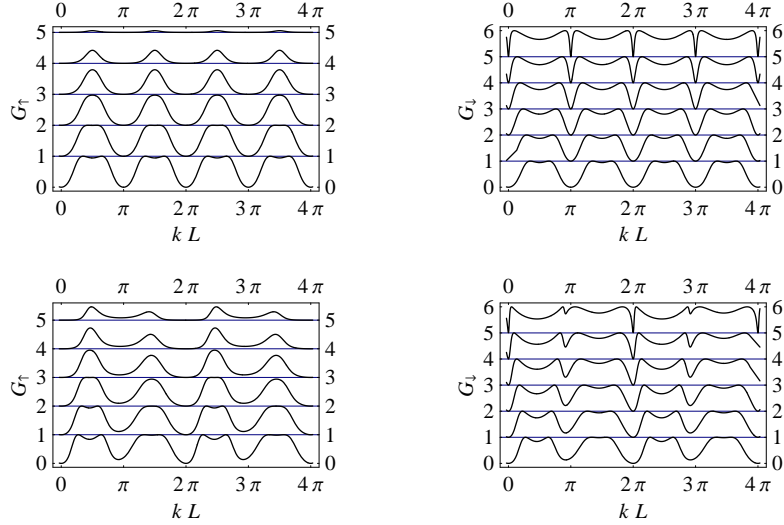}\\
\caption[Fig.4]{The spin resolved conductance $G_\sigma$ versus $k
L$ for different values of the Aharonov-Bohm flux
$\frac{\Phi_{AB}}{2\pi}=0.04\times n$, where $n=0,1,2,3,4,5$ from
bottom to top and $z=0$ (upper panels), $z=0.6$ (lower panels).
The SOC has been fixed at $\beta=1.2$. Due to the presence of the
spin-orbit interaction spin-up and down channels have different
zero-resonances structure. }\label{fig:effect_beta_cond}
\end{figure}

\begin{figure}[htbp]
\centering
\includegraphics[scale=0.5]{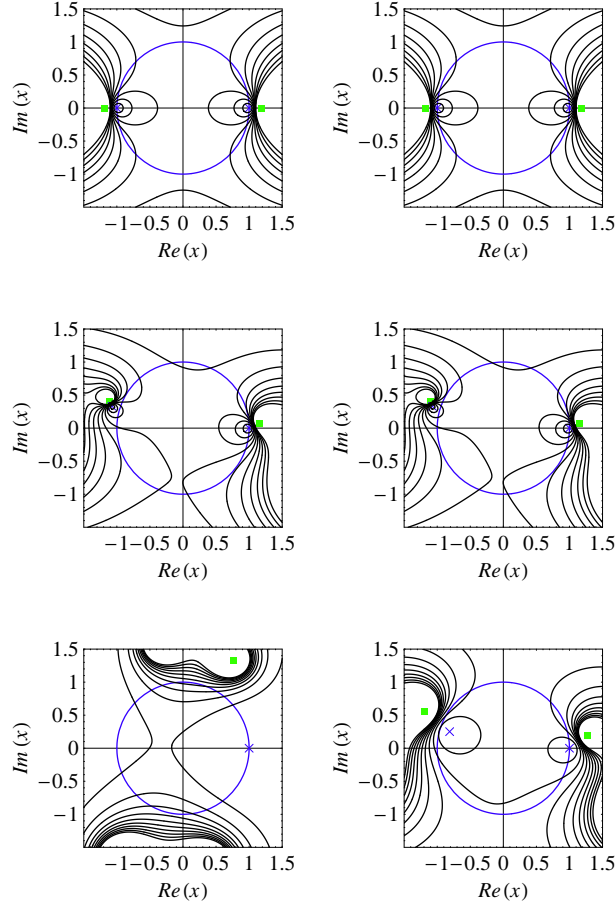}\\
\caption[Fig.7]{Zeros (denoted by a cross) and poles (denoted by a
box ) in the complex energy plane for spin up (left panel) and
spin down electrons (right panel) for the following choice of
parameters: in the upper panels $z=0$,
$\frac{\Phi_{AB}}{2\pi}=0.12$ $\beta=0$ (corresponding to curve 3
of upper panel Figure 3); in the middle panels $z=0.6$
$\frac{\Phi_{AB}}{2\pi}=0.12$, $\beta=0$ (corresponding to curve 3
of lower panel Figure 3); lower panel $z=0.6$
$\frac{\Phi_{AB}}{2\pi}=0.12$, $\beta=1.2$ (corresponding to curve
3 of lower panel Figure 4). We show the contour plot of
$|T_{\sigma}|^2$ and the unitary circle where real zeros
conductance fall. Note that the location of the zeros and poles is
on the real axis when the tunnel barrier is set to zero. The
presence of a finite tunnel barrier lift the zero-pole structure
from the real axis in the complex plane and gives rise to
asymmetric resonance lineshape in the spin-resolved conductance
(see Figure 3). The zero-pole structure is the same in both
spin-channels in absence of spin-orbit interaction while
spin-channels are distinct for $\beta$ nonzero (lower panels).}
\label{fig:effect_z_plane}
\end{figure}

To analyze the pole of the transmission
amplitude, we need to solve a fourth order algebraic equation
\begin{eqnarray}
\label{eq:fourth-order}
\mathcal{D}(x)=x^{4}(z-2i)e^{i\Phi^{\sigma}}+x^{3}(2ze^{i\Phi^{\sigma}})+2ix^{2}(4e^{i2\Phi^{\sigma}}+2e^{i\Phi^{\sigma}}+4)+x(6z
e^{i\Phi^{\sigma}})-9(z+2i)e^{i\Phi^{\sigma}}.
\end{eqnarray}
%In the limit $z\rightarrow 0$ the poles are given by
%$x_n=\sqrt{3}e^{in\pi/4}$ for $n$-odd-integer. In the opposite
In the limit $z\rightarrow \infty$, $\mathcal{D}(x) \sim z e^{i
\Phi^\sigma}(x-1)(x+3)(x-\sqrt{3}i)(x+\sqrt{3}i)$. The pole
position is independent from the spin and the transmission
amplitude is identical in both spin channels. In particular, the
pole $x=-3$, which corresponds to $k_R L=\pi$, has a large
imaginary part and gives rise to a large width of the resonance.
The effect of $z$ and of the total effective flux on the formation
of zero-resonances in the spin-resolved conductance is shown in
Figs.\ref{fig:effect_z_cond}-\ref{fig:effect_beta_cond} and in the
complex energy plane in Fig.\ref{fig:effect_z_plane}. Note that
when $z\ne 0$ structural zeros due to the scattering at the tunnel
barrier start to appear (e.g. see down panel
Fig.\ref{fig:effect_z_cond}). The last give rise to nonsymmetric
maxima in the conductance.

\subsection{spin filtering without magnets}

\begin{figure}[htbp]
\centering
\includegraphics[scale=.5]{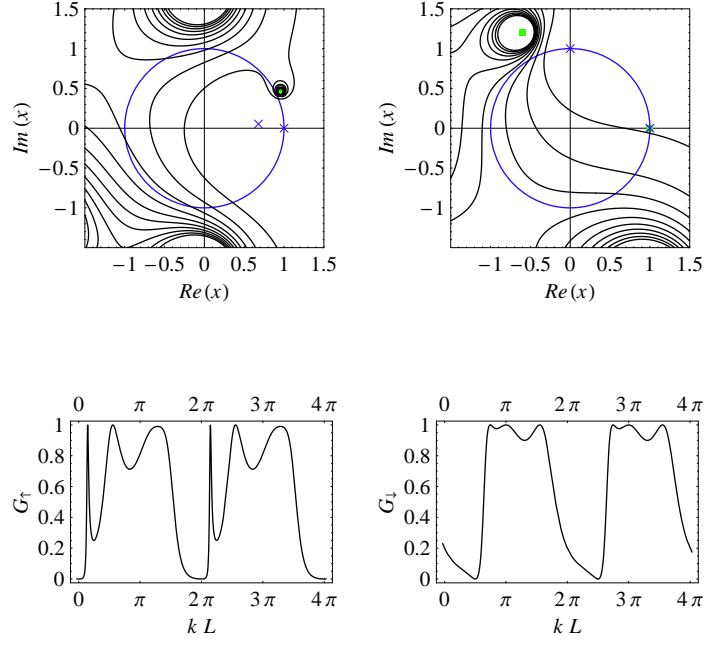}\\
\caption[Fig.10]{Conductance curves (lower panels) and zero-pole
structure (upper panels) for the following choice of parameters:
$\beta=1.2$, $\overline{k} L=\pi/2$,
$z=4\tan({\pi\over2}-{\overline{k} L\over 2 })$, ${\Phi_{AB}
\over2\pi}={1\over2}+{\sqrt{1+\beta^2}\over2}$. In the upper
panels crosses denote the location of the zeros while the squares
the location of the poles: The interference zeros in the down
spin-channel are compensated by a pole at the same position, while
the location of the zeros of second type on the unitary
circumference is determined by the value of $z$. In the lower
panels, at the zeros at $\bar{k}L=\pi/2+2n\pi$ in the down
spin-channel correspond a finite conductance in the up
spin-channel when $z$ meets the condition for spin-filtering
(\ref{eq:cond_spin_filter}), as explained in the text. Sharp
Fano-like resonances appear in the up spin channel due to the
presence of a pole with a small imaginary part.}
\label{fig:spin_filt_1}
\end{figure}

\begin{figure}[htbp]
\centering
\includegraphics[scale=.5]{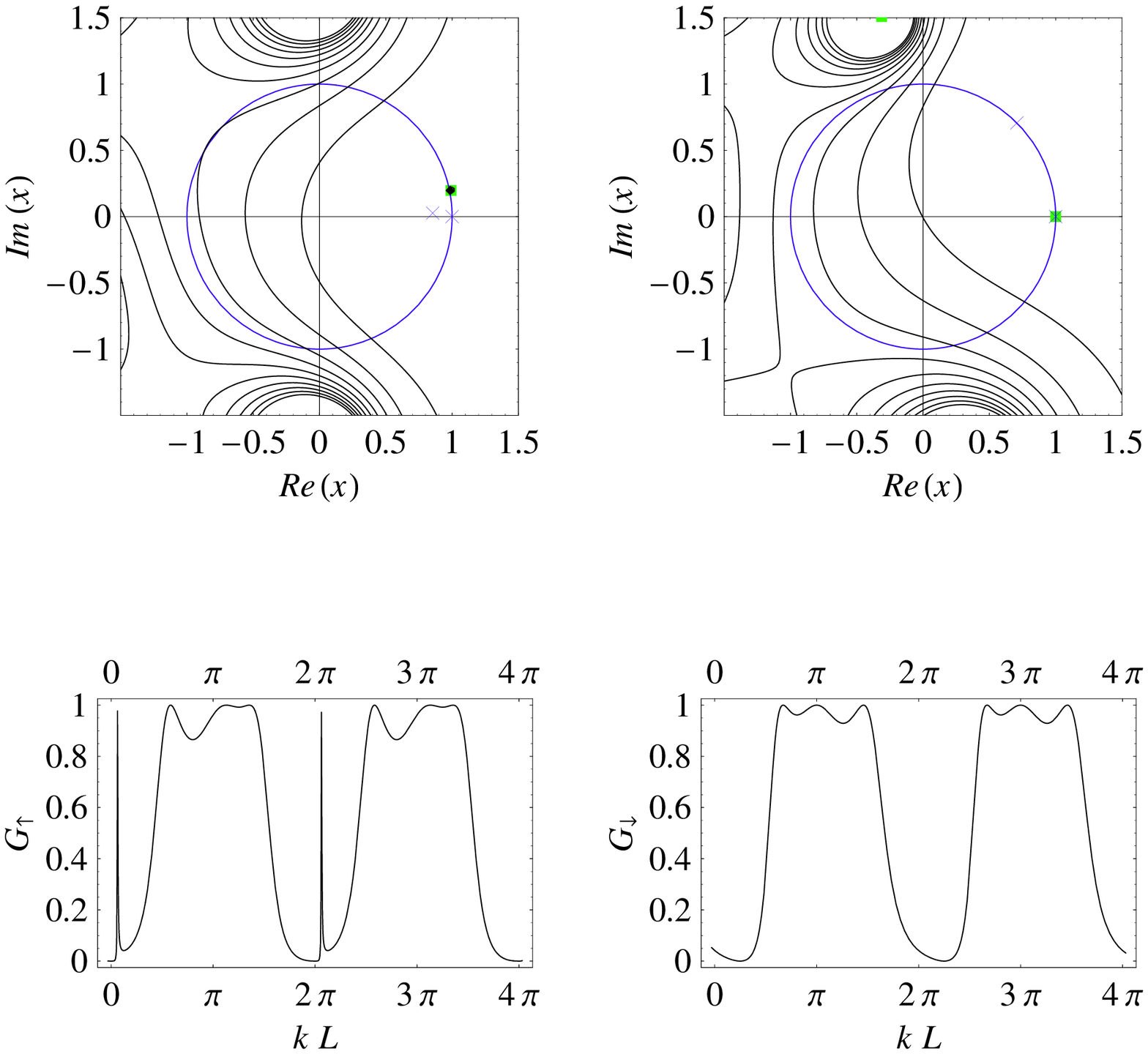}\\
\caption{The same as in Fig.\ref{fig:spin_filt_1}, with
$\overline{k} L={\pi\over4}$.} \label{fig:spin_filt_2}
\end{figure}

Hereafter we show that the tunable barrier in one arm of the ring
remarkably permits the AB ring in presence of SO interaction to
operate as a spin-selective device. As discussed above, resonances
(or poles) in the transmission amplitude do not necessarily give
rise to zeros in the conductance. This is the case when zeros and
poles compensate each other and yield a finite value of the
conductance. This property can be used to have a finite
conductance in one spin-channel while being zero in the other
channel. By examination of the transmission amplitude in the
complex energy plane, we find that the conductance zeros and
transmission resonances can be controlled by the height of the
barrier $z$ and the value of the total effective flux
$\Phi^\sigma$. In particular, by Eq.(\ref{eq:fourth-order}), we
obtain that the zeros of first kind $\phi_{0,1}$ in one
spin-channel ($x=1$) can be compensated by imposing the condition
on the denominator of the transmission amplitude
$\mathcal{D}(x_p=1)=0$, while the zeros of second kind
$\phi_{0,2}$ can be controlled by $z$. The equation
$\mathcal{D}(x_p=1)=0$ can be rewritten as
$(e^{i\Phi^\sigma}-1)^2=0$, and is solved for integer values of
the effective flux, $\Phi^\sigma=2 \pi n, n\in$Z. Such solutions
do not depend on $z$, on the contrary the location of zeros of
second type can be controlled by $z$ and correspond to energies:
\begin{eqnarray}
\label{eq:cond_spin_filter} \bar{k}L=(2 n+1
)\pi-2\arctan({z\over4}),
\end{eqnarray}
where $\bar{k}$ is the momentum of the electron for a given spin
channel with vanishing transmission amplitude. The possibility of
compensating the zeros of first kind in a selected spin channel
and controlling of the location of zeros of second type by varying
$z$ points to the use of the AB-AC ring as a spin-filter. The
spin-filtering effect is shown in Figs.\ref{fig:spin_filt_1} and
\ref{fig:spin_filt_2} for the choice $\beta=1.2$ and
$\bar{k}L=\frac{\pi}{2},\frac{\pi}{4}$, respectively, while the
other parameters are fixed according to
Eq.(\ref{eq:cond_spin_filter}). As shown in
Fig.\ref{fig:spin_filt_1} at $\bar{k}L=\pi/2+2n\pi$ the
transmission probability of electrons with spin down
$G_\downarrow$ is zero while $G_\uparrow$ is non-zero at
$\bar{k}L$. It is worth to note that in the procedure above we
have compensated the zeros of first kind only in one spin channel.
If we would like to compensate a zero of the conductance in both
spin channels we need to fix the values of the magnetic flux at
$\Phi_{AB}=(2n+1)\pi-\sigma\pi \sqrt{1+\beta^2}$, being $n \in Z$,
which is satisfied by the values of $\beta=\sqrt{4m^2-1}$, $m\in$
Z.  In this case we have a switching-effect controlled by $z$ in
both spin-channels.

To elucidate our spin-filter design we may turn to momentum
resolved tunneling properties. It is well known that spin-orbit or
Rashba interaction creates spin-orbit splitting of the conduction
electron band in the ring. The variation of the tunnel barrier
strength induces crossing points in the momentum dispersion curve
of the external leads and of the Rashba-spin-split dispersion
curve of the ring, simultaneously allowing tunneling for, e.g.
spin-up right and left-movers at one node, while tunneling of
spin-down electron is suppressed at the same node. Then electrons
ending up in opposite external leads will have opposite spins. The
tunnel barrier determines the location of the crossing points in
the energy spectrum and thus acts as a spin filter. Equivalently,
we may say that the transmission probability in the spin channel
opposite to the incident spin orientation  is the result of spin
precession along the ring branches due to SOC as considered in
Ref.\onlinecite{bulgakov_so}. The conductance zeros in the
opposite channel correspond to a frequency of precession which
reproduces the incident spin orientation at the junction. In our
case the frequency precession is a function of the total flux
enclosed in the ring and of the tunnel barrier strength.

In the context of spintronics\cite{spintr_1,spintr_2} nonmagnetic
spin-filters, as the one discussed here, are intriguing and,
perhaps, useful alternative.

\subsection{Temperature effects on the spin-filtering}

For the spin-filter realization we are proposing it's relevant to
evaluate the efficiency of the device at non-zero temperature.
Thus in the following we generalize our calculations at finite
temperature $T$. The conductance at finite $T$ is given by
\begin{equation}
G=-(e^2/h)\sum_{\sigma} \int_0^{\infty} dE \frac{\partial
f(E,\mu,T)}{\partial E}|T_\sigma(E,\Phi^\sigma,z)|^2,
\end{equation}
where $f$ is the Fermi distribution function, $\mu$ is the
chemical potential and $T$ the temperature. By transforming the
integral over energy to an integral over the momentum $k$, and
introducing the dimensionless variable $\xi=k/k_F$ ($k_F$ being
the Fermi momentum) we can express $G$ as:
\begin{equation}
G(T)=(e^2/h)\sum_{\sigma} \int_0^{\infty} d\xi \xi \frac{T_F}{2T}
\cosh^{-2}\left(\frac{T_F}{2T}
(\xi^2-\tilde{\mu})\right)|T_\sigma(2k_F\xi L,\Phi^\sigma,z)|^2,
\end{equation}
where $\tilde{\mu}$ is the (dimensionless) chemical potential in
units of Fermi energy $E_F$ and $T_F$ is the Fermi temperature. As
discussed in Ref.\onlinecite{molnar_ring_so},  for a InAs ring
radius of $R=0.25 \mu m$ and effective mass $m^\star=0.023$, the
corresponding Fermi temperature is $T_F=129.27$ K and
$k_F=20.5/R$.

\begin{figure}[htbp]
\centering
\includegraphics[scale=.5]{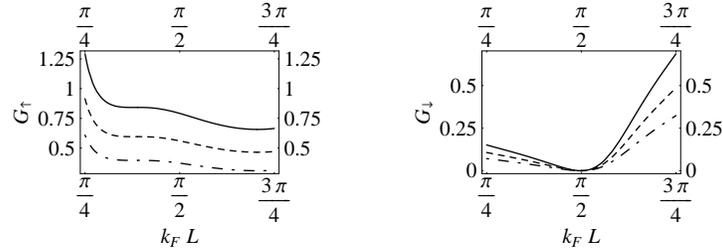}\\
\caption{Temperature dependence of spin filtering effect. The
parameters used are the same as in Fig.\ref{fig:spin_filt_1} for
$T/T_F=0.00083,0.001,0.00125$(from top to bottom) which correspond
to temperatures of the order 100 mK in InGa. Only the behavior
close to $E_F$ ($k_F L=20.5 \pi$) has been plotted (all the values
on the horizontal axis have been shifted by $-20\pi$).}
\label{fig:finite_T}
\end{figure}

In Fig.\ref{fig:finite_T} we report the results of the
spin-resolved conductance  for the following choice of parameters:
$\beta=1.2$, $\overline{k_F} L=20.5 \pi$,
$z=4\tan({\pi\over2}-{\overline{k}_F L\over 2 })$, ${\Phi_{AB}
\over2\pi}={1\over2}+{\sqrt{1+\beta^2}\over2}$ and three different
temperatures: $T/T_F=0.0008,0.0010,0.0012$. As shown, the location
of the zero in the down-spin transmission probability is not
modified by the temperature. At the zero in the transmission
probability of the down spin-channel corresponds a temperature
dependent transmission probability in the up spin-channel when $z$
meets the condition for spin-filtering
Eq.(\ref{eq:cond_spin_filter}) at fixed $\overline{k}_F$. The
value of the transmission probability in the up spin-channel is
reduced by fifty per cent when $T/T_F$ increases from 0.0008 to
0.0012, i.e. for temperatures of the order 100$mK$.  Thus at the
temperatures at which real devices are working the spin-filtering
effect is still sizeable and the efficiency remains larger than
fifty per cent.

% ----------------------------------------------------------------
\section{Conclusions}
\label{sec:concl}

In summary, we have analyzed the zero conductance resonances
appearing in an AB-AC ring as a signature of interfering resonant
states of the loop system under the influence of a magnetic flux
and of a Rashba electric field  in presence of a tunable tunnel
barrier. We have obtained, both numerically and analytically, the
explicit dependence of the transmission on the spin-orbit
coupling, the electromagnetic flux and the tunnel barrier,
elucidating the role of internal and geometrical symmetries
breaking in quantum transport and possible experimental
realizations. According to Eq.(\ref{eq:zeros_complex}) real zeros
conductance are lifted by the influence of the external fields,
being shifted into the complex plane depending on the value of the
total flux phase and of the tunnel barrier strength. In the case
of  magnetic flux (spin-orbit interaction) is the breaking of
time-reversal-symmetry (spin reversal symmetry) that destroys the
energetic degeneracy of states preventing the destruction of
interference effect at the nodes and leading to zeros.  In the
presence of a local tunnel barrier is the breaking of rotational
symmetry to lead to zeros. More importantly, we have demonstrated
how the presence of a tunable tunnel barrier can be used to
control the polarization of transmitted electrons. We have shown a
novel spin-filtering effect consisting in the possibility of
compensating the interference zeros of the transmission
probability in one spin-channel and controlling the location of
the structural zeros by tuning the tunnel barrier strength. This
effect points, once again, towards the possibility of employing
one-dimensional rings as a spin-selective devices widening the
field of usual magneto-electronics. We have also discussed the
spin-filtering effect at finite temperature and shown that it
remains sizeable when the temperature is raised up to 100 mK. We
would like to stress that our proposal is within reach with
today's technology for experiments in semiconductor
heterostructures with a two dimensional electron gas (2DEG), e.g.
InGaAs-based 2DEG, which has an internal electric field due to an
asymmetric quantum well\cite{param1,ring_exp_AC}. Indeed, spin
interference effects in Rashba-gate-controlled ring embedded in
the p-type self-assembled silicon quantum well with a quantum
point contact inserted have recently been reported\cite{ring_exp}.
The proposed spin-filter device differs by previous proposals
where the polarization of transmitted spin-polarized electrons can
be controlled via an additional inhomogeneous magnetic
field\cite{spin_filter_ring} or a symmetrically textured electric
field\cite{shen_spin_precession}. Compared to these proposals,
ours does not require additional fields and is related to a more
controllable parameter, the tunnel barrier. We would like to
stress that the presented results are valid for a one-dimensional
ring, but they can be extended to rings of finite width $w$
provided the inequality $w\ll R$ holds\cite{molnar_ring_so}.

The question wether the spin-filtering effect survives when
considering disordered-averaged quantities remains a further
interesting problem that we are taking under consideration.
Another interesting question is related to the presence of
Dresselhaus spin-orbit
interaction\cite{note_dresselhaus,dresslhaus_original,evaluation_dresselhaus_review}
that could lead to a similar conductance modulation and
spin-filtering effect.

\section{Acknowledgments}
We acknowledge enlightening discussions with S. Cojocaru.

% ----------------------------------------------------------------

%%%%%%%%%%%%%%%%%%%%%%%%Figures%%%%%%%%%%%%%%%%%%%%%%%%%%%%%%%%%%%%%%%%

\begin{thebibliography}{99}
\bibitem{note_dresselhaus} We have neglected Dresselhaus spin-orbit coupling having in mind, e.g.
InAs semiconductor where the Rashba interaction dominates.
\bibitem{note}
Generally,
$G=(e^2/h)\sum_{\sigma,\sigma'=\uparrow,\downarrow}|T_{\sigma,\sigma'}|^2$,
where $T_{\sigma,\sigma'}$ with $\sigma\ne \sigma'$ is the
spin-flip amplitude. Since the conductance does not depend on the
choice of spinor basis and we make use of the ring-spinor basis
where the spin flip amplitudes vanish, we have defined
$T_{\sigma,\sigma}=T_\sigma$.
\end{thebibliography}
\end{document}